\documentclass[10pt]{article}
\usepackage{graphicx}
\def\Title#1{\begin{center} {\Large #1 } \end{center}}
\def\Author#1{\begin{center}{ \sc #1} \end{center}}
\def\Address#1{\begin{center}{ \it #1} \end{center}}

\newcommand\pubblock{\rightline{\begin{tabular}{l} Proceedings of the CTD/WIT 2019\\ \pubnumber\\
         \pubdate  \end{tabular}}}

\newenvironment{Abstract}{\begin{quotation} \begin{center} 
             \large ABSTRACT \end{center}\bigskip 
      \begin{center}\begin{large}}{\end{large}\end{center} \end{quotation}}

\newenvironment{Presented}{\begin{quotation} \begin{center} 
             PRESENTED AT\end{center}\bigskip 
      \begin{center}\begin{large}}{\end{large}\end{center} \end{quotation}}

\def\beq{\begin{equation}}
\def\eeq#1{\label{#1}\end{equation}}
\def\eeqn{\end{equation}}

\def\beqa{\begin{eqnarray}}
\def\eeqa#1{\label{#1}\end{eqnarray}}
\def\eeqan{\end{eqnarray}}

\let\bar=\overbar

\def\Dslash{\not{\hbox{\kern-4pt $D$}}}
\def\dslash{\not{\hbox{\kern-2pt $\del$}}}

\def\msb{{\bar{\ssstyle M \kern -1pt S}}}

\usepackage{color}
\usepackage{lineno}
\usepackage{subfig}
\usepackage{hyperref}
\usepackage{amsmath}

\newcommand\pubnumber{ PROC-CTD19-103 }

\newcommand\pubdate{\today}

\def\affiliation{
Wigner RCP, Budapest, Hungary}

\def\support{\footnote{Work supported by the National Research, Development and Innovation Office of Hungary (K~128786)}}

\begin{document}


\large
\begin{titlepage}
\pubblock

\vfill
\Title{Another approach to track reconstruction: cluster analysis}
\vfill

\Author{Ferenc Sikl\'er \support}
\Address{\affiliation}
\vfill

\begin{Abstract}
A novel combination of data analysis techniques is proposed for the
reconstruction of all tracks of primary charged particles, as well as of
daughters of displaced vertices (decays, photon conversions, nuclear
interactions), created in high energy collisions. Instead of performing a
classical trajectory building or an image transformation, an efficient use of
both local and global information is undertaken while keeping competing choices
open.
The measured hits of adjacent tracking layers are clustered first with the help
of a mutual nearest neighbor search in the angular distance. The resulted
chains of connected hits are used as initial clusters and as input for cluster
analysis algorithms, such as the robust $k$-medians clustering. This latter
proceeds by alternating between the hit-to-track assignment and the track-fit
update steps, until convergence. The calculation of the hit-to-track distance
and that of the track-fit $\chi^2$ is performed through the global covariance
of the measured hits. The clustering is complemented with elements from a more
sophisticated Metropolis--Hastings MCMC algorithm, with the possibility of
adding new track hypotheses or removing unnecessary ones.
Simplified but realistic models of today's silicon trackers, including the
relevant physics processes, are employed to test and study the performance
(efficiency, purity) of the proposed method as a function of the particle
multiplicity in the collision event.
\end{Abstract}

\vfill

\begin{Presented}
Connecting the Dots and Workshop on Intelligent Trackers (CTD/WIT 2019)\\
Instituto de F\'isica Corpuscular (IFIC), Valencia, Spain\\ 
April 2-5, 2019
\end{Presented}
\vfill
\end{titlepage}
\def\thefootnote{\fnsymbol{footnote}}
\setcounter{footnote}{0}
%

\normalsize 


\section{Introduction}

The reconstruction of charged particles, of their trajectories, is an active
area of research in high energy particle and nuclear physics. The task is
usually computationally difficult (NP-hard).
Detectors at today's particle colliders mostly employ large surface
silicon-based tracking devices which sample the trajectory of the emitted
charged particles at several locations. When a charged particle crosses the
semiconducting material, it deposits energy and creates a hit by exciting
electrons to the valence band producing electron-hole pairs. The electrons or
holes, or both, are transported with an applied electric field, and their
charge is read out, amplified, and digitized.
 
The silicon-based trackers are highly segmented, they consist of several
millions of tiny pixels (dimensions of $\sim$100~$\mu$m) and of narrow but long
strips ($\sim$10~cm in length). In a high energy collision event, several
thousands of pixel and strip hits are created. Our task is to solve a
mathematical puzzle: the goal is to identify particle trajectories by
associating most of these hits to a limited number of true trajectories. The
default solution for this problem is the combinatorial track finding and
fitting~\cite{fruhwirth:1987fm} via the Kalman filter~\cite{kalman1960new}.
On the one hand, classical trajectory building utilizes mostly local
information by extending the trajectory and picking up compatible hits. On the
other hand, image transformation methods (e.g. variants of the Hough
transform~\cite{hough1962method}) collect global information on the parameters
of potential track candidates~\cite{Sikler:2017}. In the following, elements of
an alternative track reconstruction method are outlined, with the aim of
efficiently using both local and global information at the same time.

One of the goals of this study is to develop a reasonably efficient
reconstruction method for (converted) photons, this way paving the way for a
potential two-photon Bose--Einstein correlation measurement at LHC energies.
Such study needs at least two reconstructed photons in a collision event, hence
good efficiency is required. Most of the emitted photons are at low momentum,
thus their identification through calorimetry is hopeless. Fortunately, in
today's silicon trackers the chance of photon conversion is on average in the
range 60-80\%. Since a calorimetry-based track seeding for the conversion
products ($\mathrm{e^+}$, $\mathrm{e^-}$) would not work for the above reasons,
the task is to develop a track reconstruction method that works well for
detached photon conversions.

\begin{figure}[!b]
 \centering
 \includegraphics[width=0.8\textwidth]{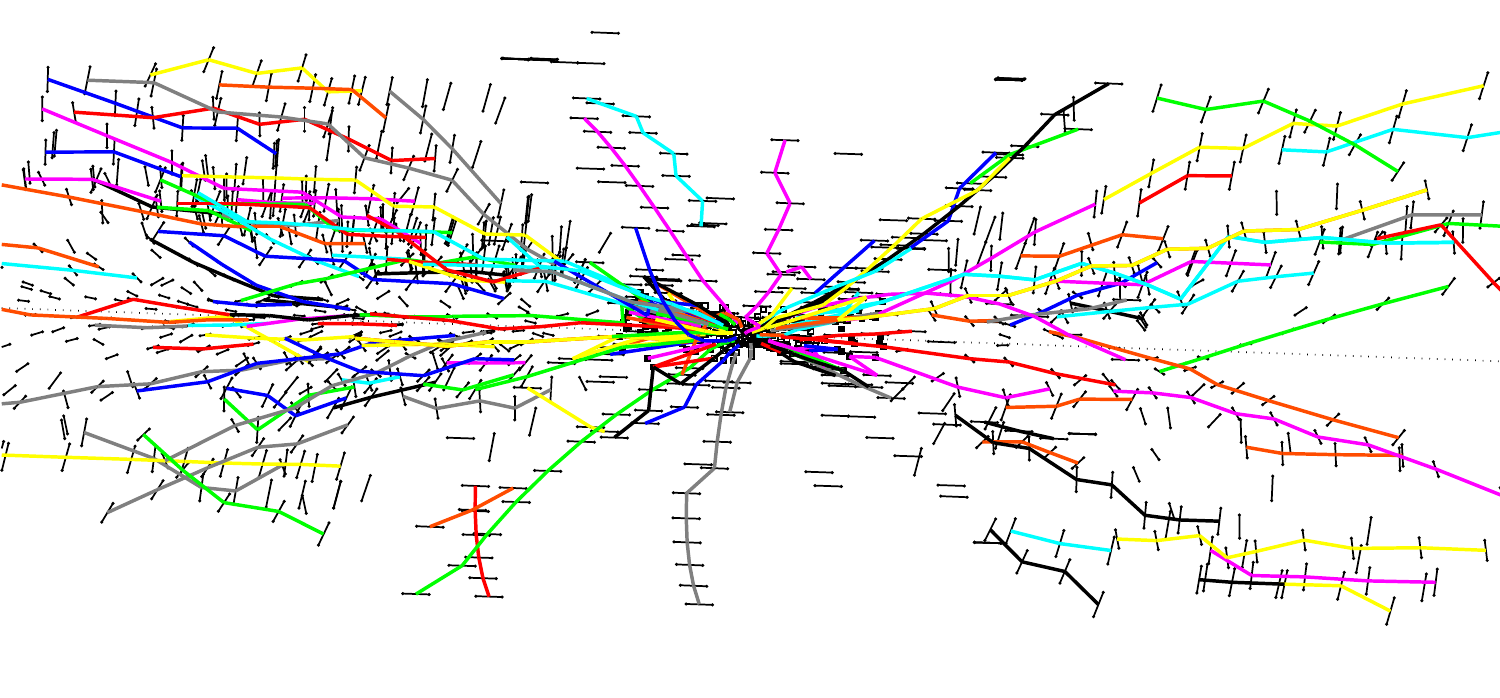}

 \caption{Chains of connected hits, taken as initial clusters in the
$k$-medians clustering method.}

 \label{fig:chains}

\end{figure}

\section{Methods}
\label{sec:methods}

The $k$-medians clustering is a robust classification
method~\cite{Steinhaus:1957,MacQueen:1967}. It aims to partition the
observations into $k$ clusters where each observation belongs to the cluster
with the nearest center. In our case the observations are the pixel or strip
hits, and the centers are the track candidates with parameters $(\eta,
q/p_\text{T}, \phi_0, z_0, r_c)$, where $\eta$ is the pseudorapidity, $q$ is
the electric charge, $p_\text{T}$ is the momentum in the transverse plane,
$\phi_0$ is the initial azimuth angle, and $z_0$ is the longitudinal, while
$r_c$ is the radial coordinate of the emission point. The method consists of
two alternating steps. First, each hit is assigned to the closest track
candidate, and then the parameters of the track candidates are updated by
refitting their associated hits to an analytic model. The process is stopped if
there are no hits changing their association (convergence) or if the number of
steps exceeds a given limit.

\begin{figure}
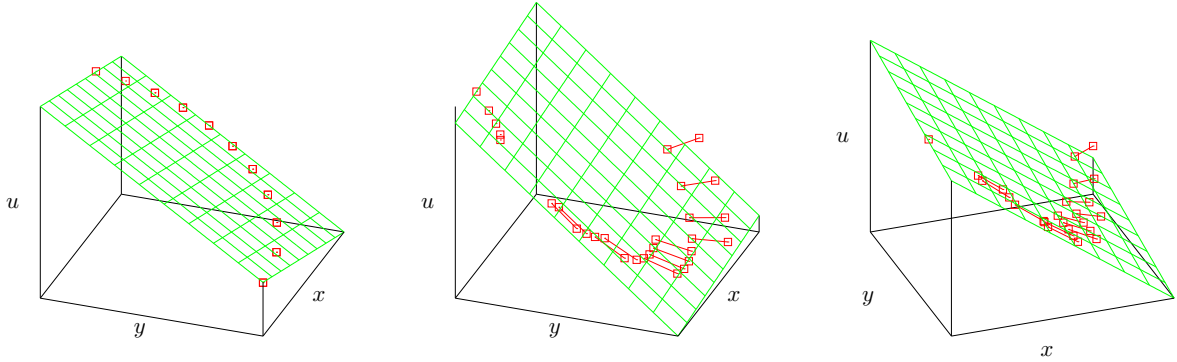


 \vspace{-0.05\textheight}
 \begin{center}
  \hspace{-0.12\textwidth}
  \resizebox{0.44\textwidth}{!}{\input{rie0}}
  \hspace{-0.12\textwidth}
  \resizebox{0.44\textwidth}{!}{\input{rie1}}
  \hspace{-0.12\textwidth}
  \resizebox{0.44\textwidth}{!}{\input{rie2}}
 \end{center}

 \vspace{-0.05\textheight}
 \caption{Transformed hits (red squares and segments) and the fitted plane
(green) corresponding to the Riemann-type fit.}

 \label{fig:riemann}

\end{figure}

It is important to choose a suitable measure of proximity. Because of outlier
hits, the use of the sum of normalized hit-to-track distances (instead of the
ordinary $\chi^2$) provides a more robust method. In our implementation, the
normalized distances are calculated through the global covariance of the
measured hits, this way, no classical trajectory building through the Kalman
filter is needed. This approach requires an analytic but precise description of
the main physical processes, such as multiple scattering, continuous energy
loss, and bremsstrahlung, with conversion to electron-positron pairs for
photons~\cite{Tanabashi:2018oca}.
In addition, non-matched candidate hits are punished by a $-2\ln(1 -
\varepsilon_\text{eff})$ terms, where $\varepsilon_\text{eff}$ is the hit
reconstruction efficiency. Non-matched measured hits come with a $-2\ln
p_\text{noise}$ contribution to the global $\chi^2$, where $p_\text{noise}$ the
frequency of reconstructed noise hits.

The locations of trajectory hits are obviously highly correlated. The
covariance between hits in layers $i$ and $j$ decays roughly proportionally to
$\rho^{-|i-j|}$, where $\rho \approx 0.8-0.9$. With that approximation, the
inverse of the covariance matrix (in the example below with four hits) is

\begin{multline*}
 V^{-1} =
 \begin{pmatrix}
  \sigma_1^2 & \textcolor[gray]{0.25}{\rho \sigma_1 \sigma_2} & \textcolor[gray]{0.50}{\rho^2 \sigma_1 \sigma_3}
                                                              & \textcolor[gray]{0.75}{\rho^2 \sigma_1 \sigma_4} \\
  \textcolor[gray]{0.25}{\rho   \sigma_1 \sigma_2} & \sigma_2^2 & \textcolor[gray]{0.25}{\rho \sigma_2 \sigma_3}
                                                                & \textcolor[gray]{0.50}{\rho \sigma_2 \sigma_4} \\
  \textcolor[gray]{0.50}{\rho^2 \sigma_1 \sigma_3} & \textcolor[gray]{0.25}{\rho \sigma_2 \sigma_3} & \sigma_3^2
                                                                & \textcolor[gray]{0.25}{\rho \sigma_3 \sigma_4} \\
  \textcolor[gray]{0.75}{\rho^3 \sigma_1 \sigma_4} & \textcolor[gray]{0.50}{\rho^2 \sigma_2 \sigma_4}
                                                   & \textcolor[gray]{0.25}{\rho   \sigma_3 \sigma_4} & \sigma_4^2
 \end{pmatrix}^{-1} = \\ =
 \frac{1}{1 - \rho^2}
 \begin{pmatrix}
  1/\sigma_1^2 & -\rho/(\sigma_1 \sigma_2) & \textcolor[gray]{0.50}{0} & \textcolor[gray]{0.50}{0} \\
  -\rho/(\sigma_1 \sigma_2) & (1+\rho^2)/\sigma_2^2 & -\rho/( \sigma_2 \sigma_3) & \textcolor[gray]{0.50}{0} \\
  \textcolor[gray]{0.50}{0} & -\rho/(\sigma_2 \sigma_3) & (1+\rho^2)/\sigma_3^2 & -\rho/(\sigma_3 \sigma_4) \\
  \textcolor[gray]{0.50}{0} & \textcolor[gray]{0.50}{0} & -\rho/(\sigma_3
\sigma_4) & 1/\sigma_4^2
 \end{pmatrix}.
\end{multline*}

\noindent
As can be seen, the inverse is tridiagonal and in the calculation of the
goodness-of-fit measure ($\sum x^T V^{-1} x$) only the differences between hits
on neighboring layers have to be taken into account. 
Track fit to the associated hits is best accomplished by the downhill simplex
method of Nelder and Mead~\cite{nelder1965}. It employs no function derivatives
but only function evaluations at the vertices of a simplex, in our case a
5-simplex.

The choice for initial clusters (tracks) is an important one. The initial
tracks could be chosen randomly, but much better performance can be achieved.
We first find all mutual nearest hit neighbors in the angular distance, with
respect to the nominal interaction point (center of the detector). Then, we
take the chains of connected hits as initial clusters (Fig.~\ref{fig:chains}).

\begin{table}

 \centering

 \caption{Main characteristics of tracking detector (silicon layers) used in
the simulation.
For the barrel layers, the layer type is shown along with the radii ($r$) of
the concentric cylinders, and their longitudinal extent
($-z_\text{max}$ to $z_\text{max}$) in the beam direction.
For the endcap layers, the layer type is shown along with their $|z|$
positions, and with the inner ($r_\text{min}$) and outer radii ($r_\text{max}$)
of their disks.
 \label{tab:layers}
 }

 \begin{tabular}{lcc}
  \hline
  Barrel  & $r$ [cm] & $z_\text{max}$ [cm] \\
  \hline
   pixels & 4, 7, 10       & 25 \\
   strips & 20, 30, 40, 50 & 55 \\
   strips & 60, 70, 80, 90, 100, 110 & 55 \\
  \hline
  Endcap  & $|z|$ [cm] & $r_\text{min}$--$r_\text{max}$ [cm] \\
  \hline
   pixels &  35,  45      &  5--15 \\
   strips &  75,  90, 105 & 20--50 \\
   strips & 125, 140, 155 & 20--110 \\
   strips & 170, 185, 200 & 30--110 \\
   strips & 220, 245      & 40--110 \\
   strips & 270           & 50--110 \\
   \hline
 \end{tabular}

\end{table}

The initial estimate of track parameters is obtained through a robust helix fit.
First, circles are found in the bending plane after centering and scaling the
hits in a given cluster. It is followed by a projection to the Riemann
sphere~\cite{riemann}, and finally planes are fitted to triplet permutations of
the projected hits. In order to allow for pixel-less tracks, strip hits are
also employed by using a few (4) representative internal points of their
segments (Fig.~\ref{fig:riemann}). During the process, we prefer a nearly equal
number of (end)points on both sides of the fitted plane, while minimizing the
sum of hit-to-plane distances.

\begin{figure}[!h]

 \centering
 \includegraphics[width=0.8\textwidth]{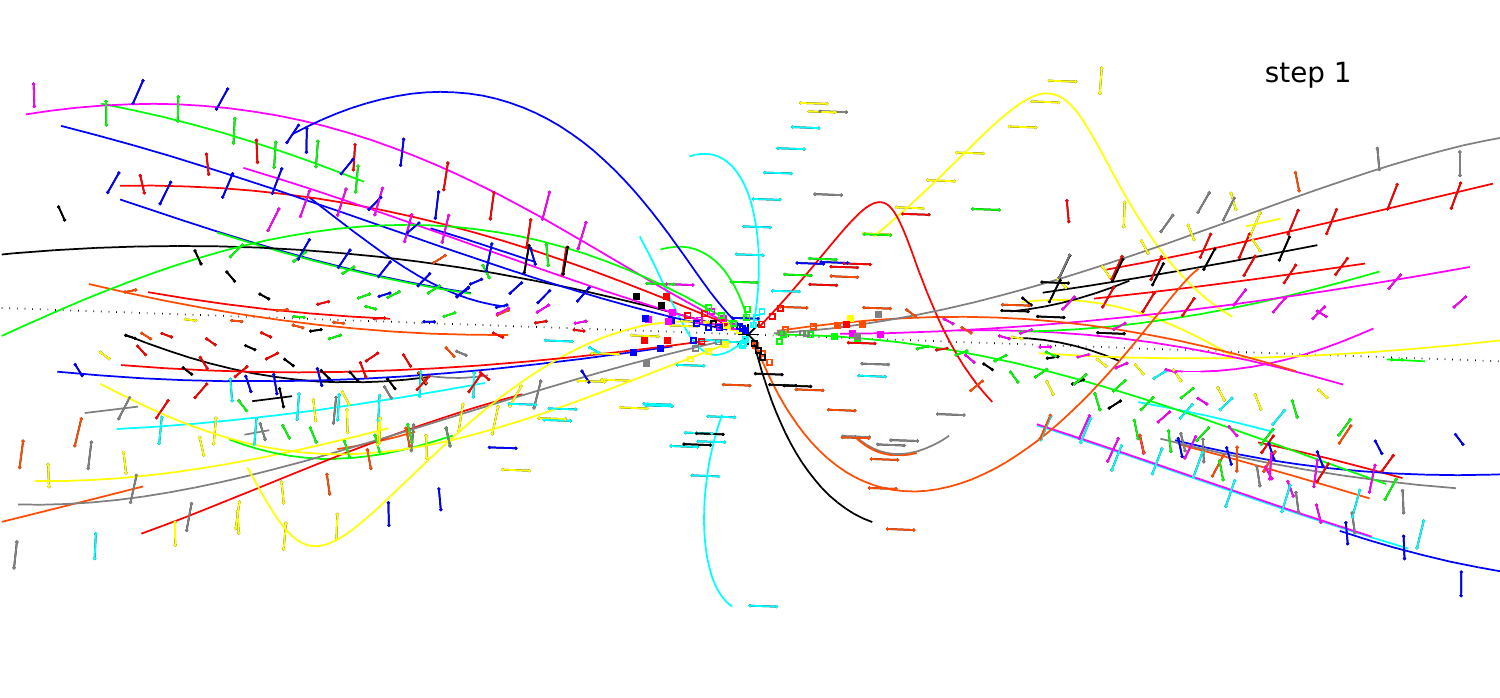}

 \vspace{-0.05\textheight}
 \includegraphics[width=0.8\textwidth]{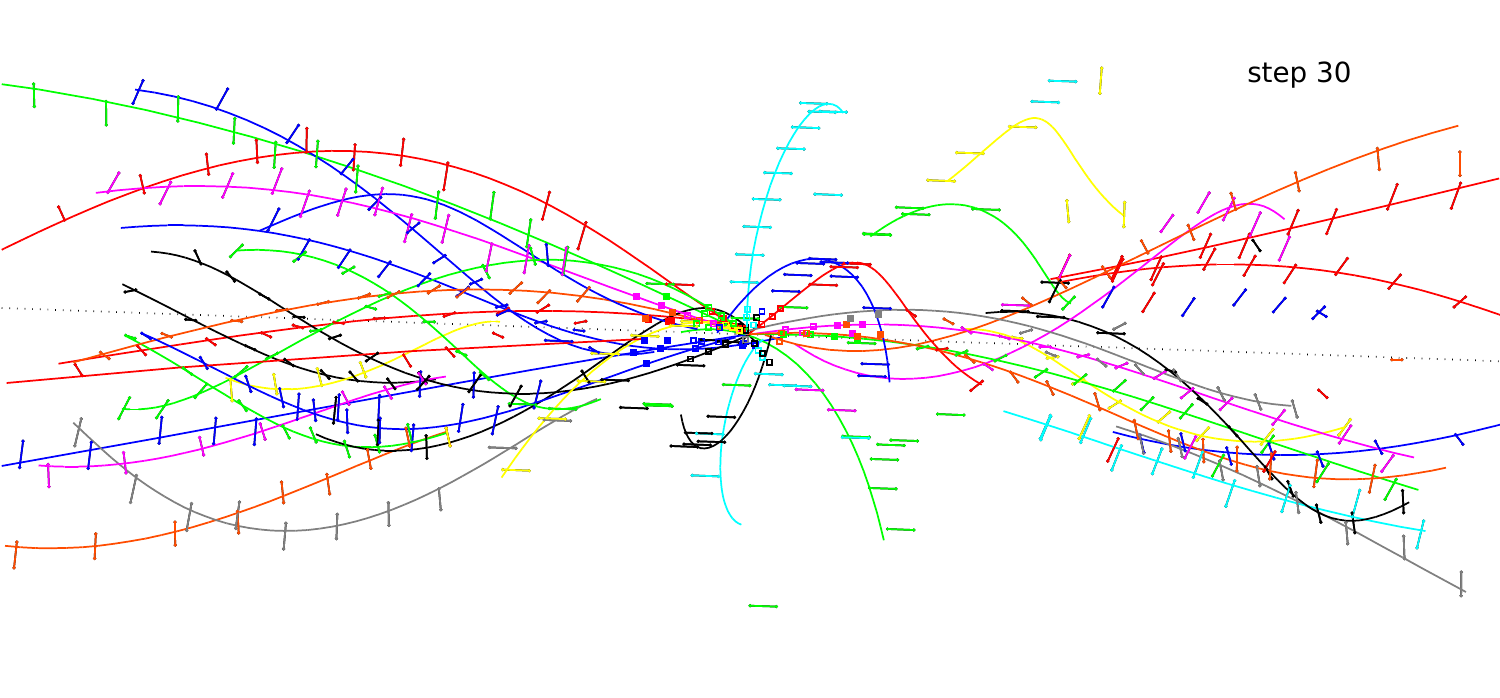}

 \caption{Hits and track candidates, their trajectories (colored curves), after
the first iteration (top), and after the 30th iteration (bottom). The event is
identical to the one displayed in Fig.~\ref{fig:chains}.}

 \label{fig:steps}

\end{figure}
 
\section{Simulation results}

The above ideas are demonstrated on a simplified detector model, with
cylindrical and disk-type layers of pixel and strip silicon sensors, in a
barrel-and-endcap layout (Table~\ref{tab:layers}). The thickness of the pixel
layers is 2\% in radiation length units ($X_0$), while for strip layers it is
set to 5\%. The tracker detector is immersed in a homogeneous magnetic field of
$B_z =$ 3~T, where $z$ is in the beam direction.
Altogether thousand collision events with 24, 48, or 96 primary charged
particles, and half as many converted photons, are generated. The primary
interaction points are chosen on the $z$-axis, according to a normal
distribution with a standard deviation of $\sigma_z =$ 5~cm.

\begin{figure}[!b]
 \centering

 \vspace{-0.075\textheight}
 \includegraphics[width=0.8\textwidth]{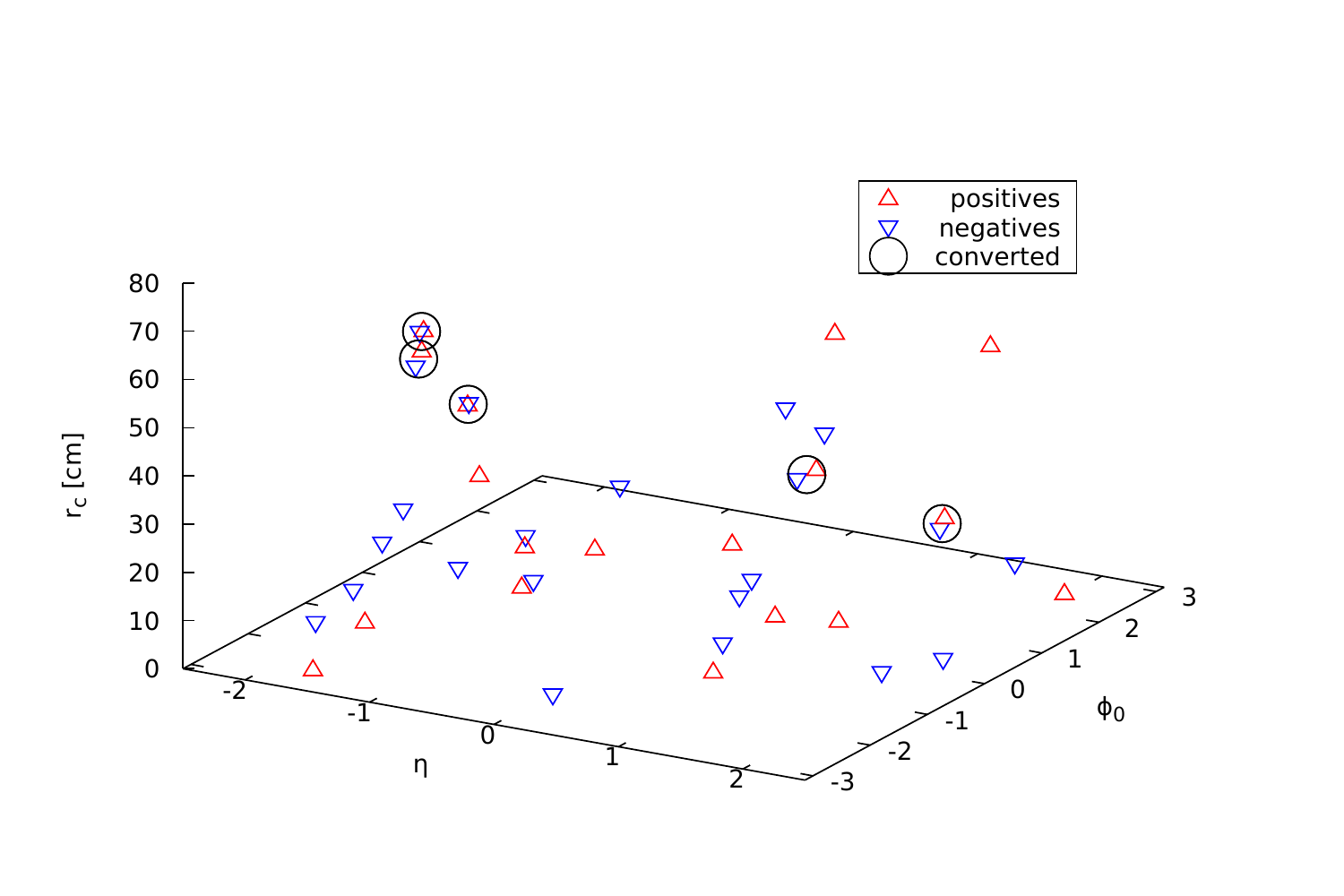}

 \vspace{-0.075\textheight}
 \includegraphics[width=0.8\textwidth]{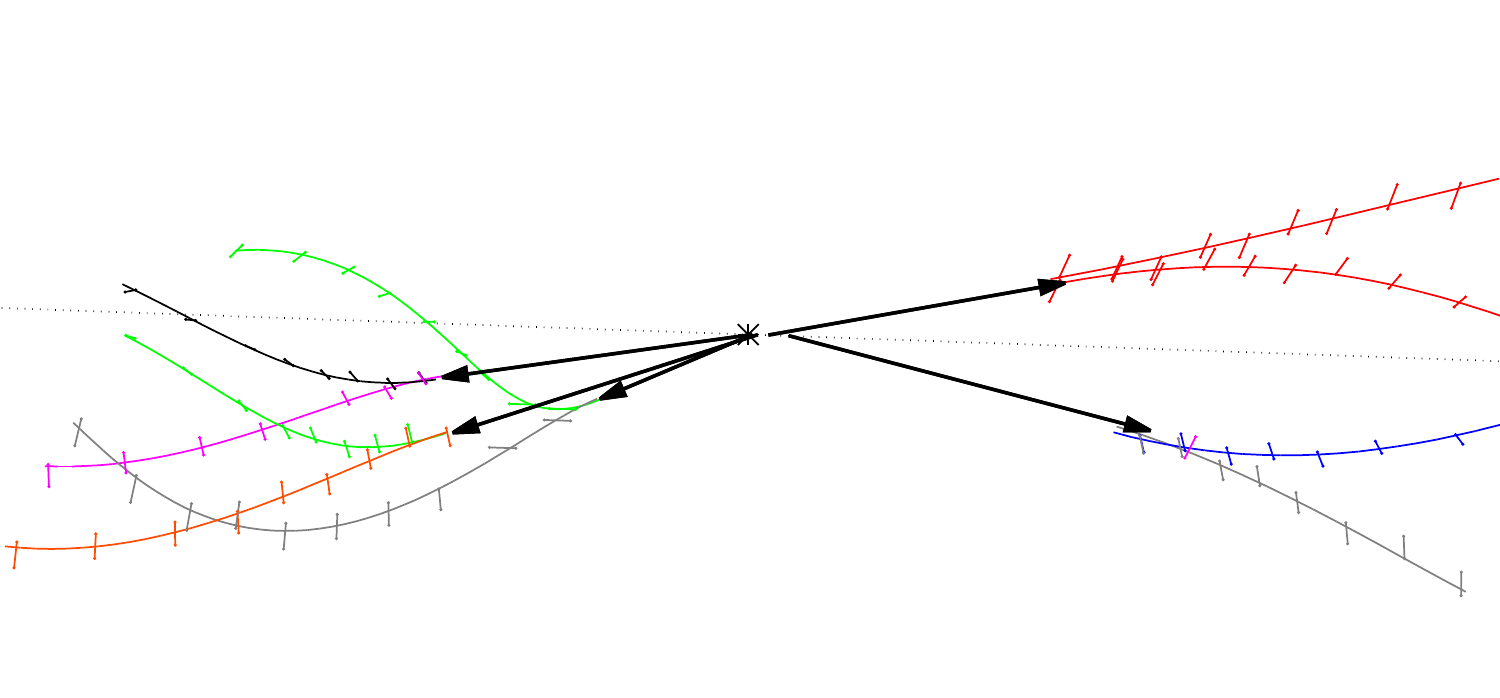}

 \caption{Top: identification of photon conversions in the $(\eta, \phi_0, rc)$
space of track candidates. Bottom: hits and track candidates, their
trajectories, corresponding to the electron and positron tracks (colored curves)
coming from photon conversions (thick black arrows). The event is identical to
the one displayed in Fig.~\ref{fig:steps}.}

 \label{fig:conversions}

\end{figure}

The generated charged particles have a uniform distribution in pseudorapidity
in the range $-2.5 < \eta < 2.5$ and in azimuthal angle $\phi$. Their
$p_\text{T}$ distribution is proportional to $p_\text{T}^2
\exp(-p_\text{T}/p_0)$, where $p_0$ is chosen to be 0.2~GeV/$c$.
Photons are generated with similar $\eta$, $\phi$, and $p_\text{T}$
distributions but with $p_0 =$ 0.1~GeV/$c$. The momentum distribution of their
conversion products (electrons and positrons) are chosen according to the
simplified Tsai's formula~\cite{Tanabashi:2018oca}.

The layer-to-layer tracking of charged particles in the homogeneous magnetic
field is performed by piecewise helices.
All relevant physical processes are properly modelled, such as multiple
scattering and specific energy loss (for all charged particles), bremsstrahlung
(for electrons and positrons), and photon conversion~\cite{Tanabashi:2018oca}.
The uncertainty of the local position measurement is modeled according to a
normal distribution with a standard deviation of 50~$\mu$m.
The efficiency of hit ($\varepsilon_\text{eff}$) reconstruction is taken to be
98\%.

\begin{figure}
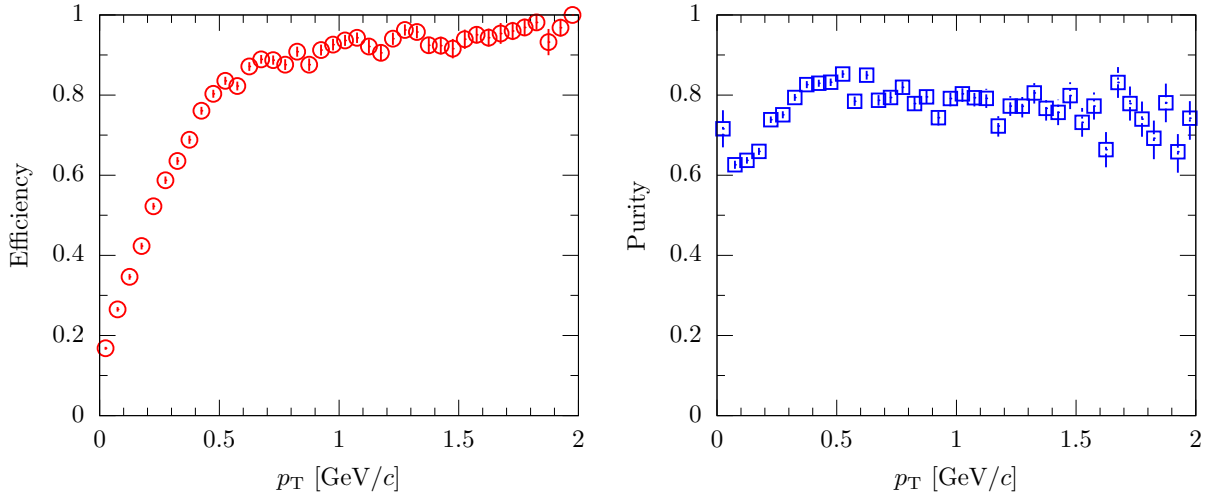

 \centering

 \resizebox{0.49\textwidth}{!}{\input{efficiency_pt} }
 \resizebox{0.49\textwidth}{!}{\input{purity_pt} }

 \caption{Efficiency (left) and purity (right) of the proposed reconstruction
method for primary charged particles.}

 \label{fig:effPur}

\end{figure}

\begin{figure}
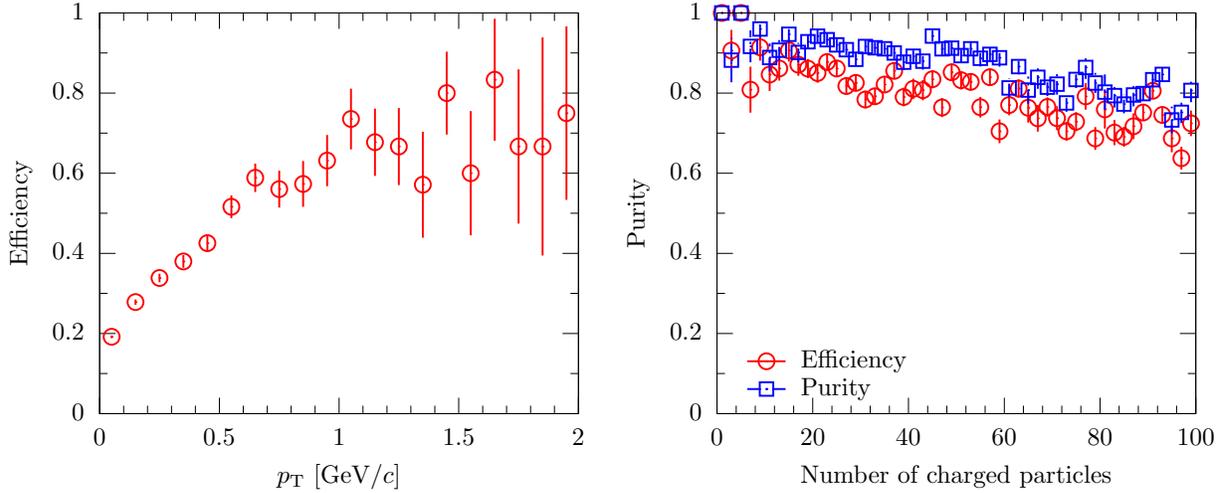

 \centering 

  \resizebox{0.49\textwidth}{!}{\input{efficiency_pt_elec} }
  \resizebox{0.49\textwidth}{!}{\input{performance_vs_multi} }

 \caption{Left: efficiency of the proposed reconstruction method for conversion
electrons. Right: multiplicity dependence of efficiency and purity for primary
charged particles.}

 \label{fig:effPur_elec_multi}

\end{figure}

As results of the track finding steps outlined in Section~\ref{sec:methods},
hits and track candidates, their trajectories, after the first and the 30th
(final) $k$-medians iterations are shown in Fig.~\ref{fig:steps}.
For primary particles the tracking efficiency in the range $p_\text{T} >$
0.5~GeV/$c$ is observed to be around of 90--95\% (Fig.~\ref{fig:effPur}). It decreases towards very low
transverse momenta and reaches 50\% near 0.2~GeV/$c$. The purity is around
80\%, independent of $p_\text{T}$.
Photon conversions are found by searching for close positively- and
negatively-charged track candidates in the $(\eta, \phi_0, rc)$ space
(Fig.~\ref{fig:conversions}-top). The corresponding electron and positron tracks are
plotted in Fig.~\ref{fig:conversions}-bottom.
For conversion electrons (and positrons), the tracking efficiency in the range
$p_\text{T} >$ 0.6~GeV/$c$ is around 70\%, with a slight decrease towards lower
transverse momenta, and it reaches 30\% near 0.2~GeV/$c$
(Fig.~\ref{fig:effPur_elec_multi}-left).

According to these simple tests, the measures mentioned above only slightly
depend on the number of primary charged particles in the studied multiplicity
range (Fig.~\ref{fig:effPur_elec_multi}-right).
The performance is further increased by using elements from a more
sophisticated Metropolis--Hastings MCMC algorithm~\cite{hastings}, namely by
sometimes adding new track hypotheses and removing unnecessary ones during the
iteration process.

\section{Conclusions}

A novel combination of data analysis techniques was proposed for the
reconstruction of all tracks of primary charged particles, as well as of
daughters of displaced vertices, created in high energy collisions. Instead of
performing a classical trajectory building or an image transformation, an
efficient use of both local and global information is undertaken while keeping
competing choices open.

The measured hits of adjacent tracking layers are clustered first with the help
of a mutual nearest neighbor search in the angular distance. The resulting
chains of connected hits are used as initial clusters and as input for a
cluster analysis algorithm, the robust $k$-medians clustering. This latter
proceeds by alternating between the hit-to-track assignment and the track-fit
update steps, until convergence. The calculation of the hit-to-track distance
and that of the track-fit $\chi^2$ is performed through the global covariance
of the measured hits. The clustering is complemented with elements from a more
sophisticated Metropolis--Hastings MCMC algorithm, with the possibility of
adding new track hypotheses or removing unnecessary ones.

Preliminary studies show that the proposed method provides reasonable
efficiency and purity for the reconstruction of converted photons, this way, it
opens the way towards an efficient identification of low momentum converted
photons.

\bibliographystyle{myStyle}
\bibliography{sikler_cluster_proc}

\end{document}